# In-situ force microscopy to investigate fracture in stretchable electronics: insights on local surface mechanics and conductivity

*Giorgio Cortelli, Luca Patruno, Tobias Cramer\*, Beatrice Fraboni, Stefano de Miranda\**

*G. Cortelli, Dr. L. Patruno, Prof. S. de Miranda*
*Department of Civil, Chemical, Environmental and Materials Engineering*
*University of Bologna*
*Viale del Risorgimento 2, 40136, Bologna, Italy*

*Prof. T. Cramer, Prof. B. Fraboni*
*Department of Physics and Astronomy*
*University of Bologna*
*Viale Berti Pichat 6/2, 40127, Bologna, Italy*

E-mail: *tobias.cramer@unibo.it*, *stefano.demiranda@unibo.it*

**Abstract**

Stretchable conductors are of crucial relevance for emerging technologies such as wearable electronics, low-invasive bioelectronic implants or soft actuators for robotics. A critical issue for their development regards the understanding of defect formation and fracture of conducting pathways during stress-strain cycles. Here we present a novel atomic force microscopy (AFM) method that provides multichannel images of surface morphology, conductivity, and elastic modulus during sample deformation. To develop the method, we investigate in detail the mechanical interactions between the AFM tip and a stretched, free-standing thin film sample. Our findings reveal the conditions to avoid artifacts related to sample bending modes or resonant excitations. As an example, we analyze strain effects in thin gold films deposited on a soft silicone substrate. Our technique allows to observe the details of microcrack opening during tensile strain and their impact on local current transport and surface mechanics. We find that although the film fractures into separate fragments, at higher strain a current transport is sustained by a tunneling mechanism. The microscopic observation of local defect formation and their correlation to local conductivity will provide novel insight to design more robust and fatigue resistant stretchable conductors.

1. **Introduction**

Stretchable conductive thin films on polymeric substrates are of pivotal importance for several novel applications such as flexible and wearable electronics, stretchable bioelectronic implants, microelectromechanical systems, or soft actuators for robotics.[1–3] In these applications, the electrical properties of the thin film have to withstand the mechanically demanding deformations occurring during device operation and wear.[4] A fundamental problem regards the mismatch in elastic properties between conductive thin film and the dielectric substrate material. Conductivity relies on rigid metals or conducting polymers whereas the substrate is made of soft elastomers to warrant device compliance. Differences in elastic moduli spanning orders of magnitude are often the case and lead to the build-up of interfacial stress during deformation. The consequence are defect formation and defect evolution as observed in the form of thin film necks, cracks, fracture, and delamination. Understanding the microscopic mechanism of defect formation as well as the impact of defects on the electric properties is of paramount importance to optimize the mechanical wear resistance as needed in future application scenarios. Despite this need, microscopy techniques that characterize local morphological, mechanical, and electric properties of the metal layers in-situ during the deformation process are still missing. [5,6] Only such multichannel imaging techniques will ultimately enable the correlation of morphological defects to the electrical response.

To date, several studies demonstrate optical or electron microscopy techniques combined with mechanical stretching to provide rapid imaging of the metallic surface during sample deformation.[1–3,7–9] Digital image analysis allows then to quantify the local strain field and to obtain quantitative information on the onset of crack formation, the crack length, and the crack density as a function of the strain.[6,10–12] These are all parameters of central importance to describe the fracture mechanics of such thin films. The drawback of optical or electronic imaging techniques comes from the reflection-based image reconstruction that cannot provide quantitative information on surface height changes. Accordingly, it is difficult to clearly distinguish through and part-through surface cracks or necking structures in tensile strain experiments or to distinguish bulged structures from delaminated ones. Instead, quantitative morphological information is provided by atomic force microscopy or confocal laser scanning microscopy. First reports demonstrate the applicability of these microscopy techniques in in-situ experiments combined with macroscopic mechanical testing and conductivity measurements.[13] Such data is highly needed to establish quantitative models for predicting the degradation of conductivity as a function of strain. [1,3,4,12,14–16]

Despite these successes, several crucial local properties of the metal thin film remain experimentally not accessible, hampering the development of more precise and realistic mechanical models. This regards in particular local conductivity, which is notoriously influenced by the development of the crack pattern. In fact, current models assume ohmic conductance in the defect-free parts of the metal layer, whereas through-thickness cracks are considered as completely isolating barriers.[17,18] Relying on these two assumptions, the determination of the conductivity of the cracked metallic film reduces to the determination of the geometry of the ohmically conducting pathway connecting through the fractured film. Once it is known, one can estimate the increase in the effective path length and its width reduction during the fracture process to predict the reduction in macroscopic conductivity. So far, no experimental confirmation has been obtained on these central assumptions. Observations such as the degradation of conductivity at large strain values point already to a more complicated role of local conductivity. [17]

To address these issues, we report here an in-situ atomic force microscopy method that provides multichannel images of local surface morphology, mechanics, and conductivity on strained metal thin films. The method employs fast repetitive force spectroscopy experiments combined with a conducting AFM probe. Its application on a free-standing strained sample is demonstrated allowing efficient acquisition of multichannel images at different strain values. Possible artifacts due to substrate bending or resonant vibration are analyzed in detail to derive the optimized experimental conditions for AFM measurements on free-standing samples. As an example, we investigate the fracture of a thin gold film deposited with an adhesion layer on silicon elastomer substrates. Similar films have important roles as stretchable conductor lines in implantable electronics.[19–22] The combination of topographic, micromechanical and electrical imaging channels in our microscopy technique allows to clearly distinguish different defect types and to understand their effect on the macroscopic properties. For example, we find that although the macroscopic conductivity shows linear ohmic conducting behavior, the local current paths connecting individual fragments behave in a strongly non-linear way at increasing strain. The observation points to the relevance of tunneling based transport processes in microcracked geometries.

## 2. Result

Our in-situ atomic force microscopy technique is based on the experimental setup shown in **Figure 1a**. This setup provides multichannel acquisitions to map the surface morphology, micro-mechanics, and local conductivity as a function of the strain applied to the sample. A conductive AFM probe is used to perform simultaneously conductive AFM and force spectroscopy. During a conductive AFM experiment, a bias voltage is applied between the AFM probe and the sample to measure the local electrical current ($I_{local}$) that enters the probe through the contact area with the thin film. At the same time, the macroscopic current ($I_{macro}$) flowing in the entire sample is measured with a SMU. Uniaxial tensile strain is applied by a custom-designed strain stage in which a screw controls the distance between two clamps that hold the free-standing sample (**Figure S1**). The multichannel images are obtained through fast repetition force spectroscopy. Each pixel of the image corresponds to a force spectroscopy performed with a conductive probe and can thus provide information on surface height, stiffness and local current measured at a threshold force.

To test the AFM method, we analyzed the strain response of a thin gold film deposited with a chromium adhesion layer on a silicon elastomer substrate (PDMS). Such films are considered a prototype of a stretchable conductors, as during strain a pattern of microcracks evolves, that absorbs the strain by 3D deformation while maintaining an inter-connected conductive pathway in the gold layer.[23] **Figure 1b** shows an optical microscopy image of such a microcracked film with a gold thickness of 18nm and PDMS substrate of 2.1mm thickness as investigated in our experiments. A typical measurement curve acquired with force spectroscopy and conductive AFM on a gold region is shown in **Figure 1c**.[24] During the measurement the AFM probe is pushed into and retracted from the sample at constant speed while the force, the current and the tip position relative to the surface (indentation) are measured. When the tip contacts the conductive film, an increase of force and a sudden rise in current are recorded. During indentation, the current reaches a saturation value while the force increases linearly. Upon retraction, the force follows the loading curve as expected for an elastic response. Due to adhesion, the surface sticks to the tip when it is displaced above the surface and a negative force is measured while the current remains stable until the contact is lost during snap off.

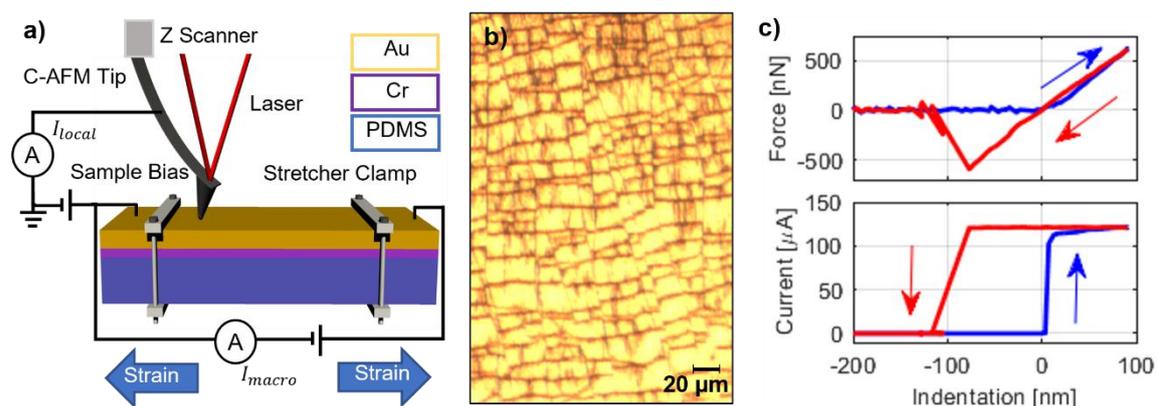

Figure 1 Experimental design of multichannel AFM in-situ experiments. a) Scheme of the experimental setup. The stretcher used to apply tensile strain to the sample is represented by its clamps. The electrical circuit permits to apply a bias between tip and sample to measure local currents entering into the conducting AFM tip and to measure also the macroscopic sample conductivity. b) Optical microscopy image of the investigated microcracked gold layer. c) Force-indentation and current-indentation curves obtained in a single-pixel acquisition. Blue and red arrows represent the load and unload, respectively.

Since the images are obtained through fast repetition of the force spectroscopy experiment, it is crucial to investigate the possible excitation of resonant oscillations of the free-standing sample that would interfere with the AFM characterization. To study resonant oscillations, we used the experimental setup shown in **Figure 2a**. Oscillation modes were excited by sound produced at different frequencies with a speaker connected to a function generator. The oscillations of the sample were probed by an AFM tip in contact with the sample. The AFM tip deflection signal was recorded with a Lock-in amplifier and analyzed as a function of frequency. **Figure 2b** shows the measured frequency response of the sample vibrations at different strain. As expected, an increase in strain corresponds to a shift of the resonance peak to higher frequencies. To ensure that the AFM measurements are not affected by the sample's vibrations, the fast repetitive acquisitions must operate at frequencies below the first resonance peak. The resonant frequencies of the sample depend on its dimensions and the strain applied. Therefore, these parameters define limits in which the setup can operate.

To understand the dependence of the resonant frequencies on the sample dimensions and the strain, we developed a model starting from the well-known Rayleigh quotient method for vibration analysis and compared it with the experimental results.[25] The sample is modeled as a pre-stressed plate clamped at two ends. Geometrical variations due to large strains and Poisson effect have been accounted for. In fact, as the strain increases, the length increases, while the width and thickness decrease. With these considerations we obtained the following formula for the first vibrating mode of a rectangular plate clamped at two ends to which a pre-stress is applied:

$$f_0^{1st}(\varepsilon) = \frac{\pi h}{3L^2} \frac{(1-\nu\varepsilon)}{(1+\varepsilon)^2} \sqrt{\frac{E}{\rho(1-\nu^2)}\left(1 + \frac{3L^2}{\pi^2 h^2} \frac{\varepsilon(1-\nu^2)(1+\varepsilon)^2}{(1-\nu\varepsilon)^2}\right)} \quad (1)$$

where $L$ and $h$ are the length and the thickness of the sample, respectively, while $\rho$ is the density, $E$ is the elastic modulus, $\nu$ is the Poisson's ratio and $\varepsilon$ is the strain. More details can be found in the Supporting Information. We then compared the experimental results with the analytical model. To do so, we considered the dimensions of our sample and the PDMS material parameters ($h = 2.1\ mm$, $L = 19.5\ mm$, $\nu = 0.5$, $\rho = 0.965\ g\ cm^{-3}$, $E = 2.21\ MPa$). The elastic modulus of our PDMS substrate was determined from indentation experiments relying on the Hertz model for the indentation of a rigid spherical tip into an elastic half-space.[26] As shown in **Figure 2c**, the analytical model agrees well with the experimental results, even though no fitting parameters are involved. To exclude the excitation of substrate resonant oscillations, the force spectroscopy experiments have to be conducted in a frequency space below the first oscillation peak. The power spectral density of tip-sample interaction modes occurring during the approach and retract movement of the tip is shown in **Figure 2d**. The inset displays the related time transient of the tip-sample force as caused by vertical movements during consecutive fast force spectroscopy measurements. The power spectral density demonstrates that significant frequency components are only below 100 Hz. This is sufficiently lower than the frequency of the first sample oscillation mode starting at 300 Hz to exclude excitation of sample oscillatory modes and to warrant stable AFM measurement conditions.

A second possible artifact during force spectroscopy measurements on free-standing samples regards global sample deflection. It must be ensured that the displacement due to bending of the free-standing sample, $\delta_{flex}$, can be neglected with respect to the local displacement under the tip, $\delta_{hertz}$, during indentation. Combining the expressions for beam deflection with the Hertz model, the ratio between the deflection and the indentation is expressed as:

$$\frac{\delta_{flex}}{\delta_{hertz}} = \frac{L^3}{192I}\left(\frac{8FR(1-\nu^2)}{9E}\right)^{1/3} \quad (2)$$

where $I = \frac{Bh^3}{12}$ is the beam cross section inertia, and $B$ is the sample width. Considering our experimental case, we obtain $\delta_{flex}/\delta_{hertz} = 0.0008$. Therefore, the bending of the sample is negligible with respect to the indentation. For this estimation we neglected the impact of the thin gold layer. A detailed discussion of **Equation 2** is provided in the Supporting Information.

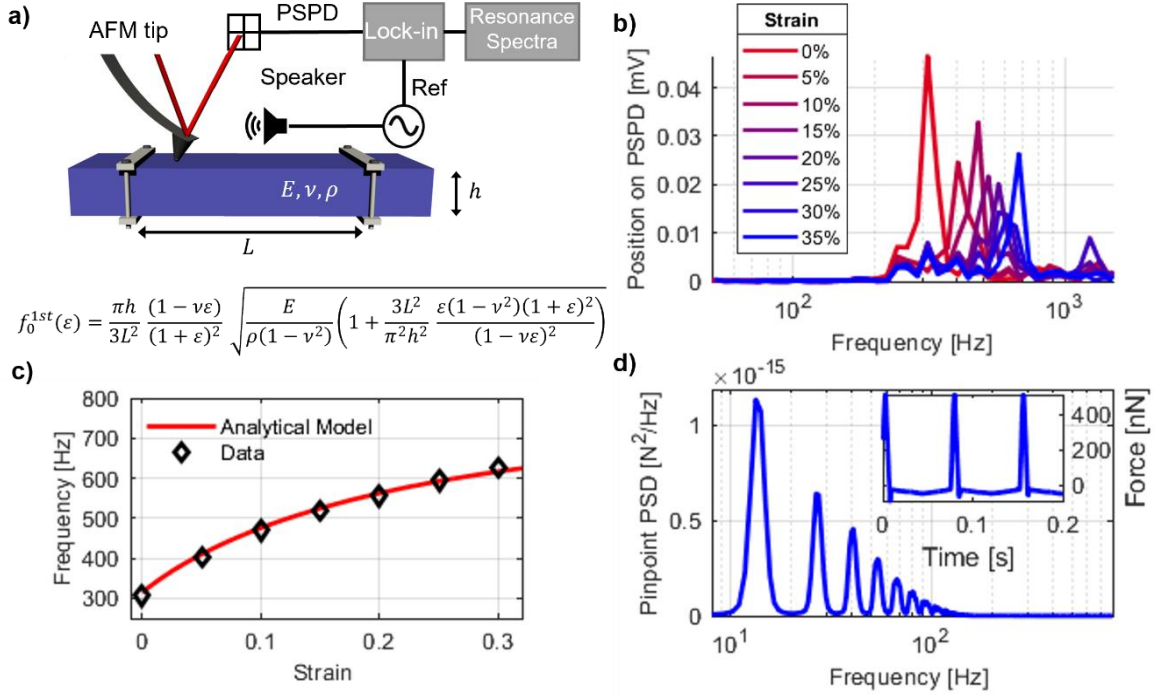

Figure 2. Investigation of measurement artifacts in AFM experiments on free standing stretched samples. a) Scheme of the experimental setup to measure the stretched sample's resonant frequencies. Oscillation modes are excited by sound produced at different frequencies. The response of the sample is measured with AFM in contact mode. b) Sample oscillation frequency spectra at different strains. Note that the peak corresponding to the first mode moves to the right as the strain increases. c) Peak frequency, extracted from the data in Figure 2b, as a function of strain. The red line corresponds to the analytical model. d) Power Spectral Density of forces between sample and AFM probe occurring during fast repetitive force spectroscopy acquisition. Note that tip-sample interactions occur at frequencies below the first vibration mode of the free-standing sample thus excluding possible resonant interactions.

Knowing the operational limits of our setup, we tested the in-situ atomic force microscopy method on the stretchable conductor prototype PDMS/Cr/Au. **Figure 3** shows the surface height, stiffness, and current maps for three different strain values, 0%, 5%, 10% respectively. The blue dashed lines indicate the position of the profile reported at the bottom ofFigure **3**. The straining direction is indicated by the red arrows. The heights of the morphology map represent the Z position when the maximum force value is reached. The stiffness is calculated as the slope of the force-indentation curve, while the local electrical current is measured at maximum force. The morphology maps (Z Height) and the extracted profiles show that at 0% strain the microcracks are closed. As the strain increases, the size of the microcracks increases. As expected, cracks are oriented in a direction normal to the strain. The maps show that the density of microcracks in the gold thin film is not strain-dependent: no new cracks are formed during the experiment. In the morphology map at strain 10% a second effect is present, the buckling of the gold layer. This is due to the Poisson effect, i.e. the compression of the sample in the direction orthogonal to that in which the strain is applied.[27] The stiffness maps show that the gold film is more compliant near the cracks. In fact, the width of the microcracks in the stiffness maps appears greater than about 1 µm, as measured from the profiles. It can be conjectured that this effect is caused by the loss of rigidity of the gold film when it is indented close to the microcracks.[24,28] Also, note that the stiffness of the gold film does not vary with strain. This indicates that the effect of strain on the single gold region is mainly to increase the distance to other regions, i.e. widen the microcracks.

From the stiffness maps, we estimated the elastic moduli of the metal thin film and the polymeric substrate. Considering the stiffness density distributions shown in **Figure 4a** one can see the presence of two peaks. The lower stiffness peak corresponds to the substrate and is therefore the result of the force spectroscopies performed in the microcracks. As the strain increases, the part of the image where microcracks are present increases. Therefore, the peak is more and more evident. On the other hand, the second peak corresponds to the force spectroscopies made on the regions with the gold film. Fitting the stiffness density distributions with a sum of two Gaussians, we estimated the average stiffness values of the PDMS (dotted lines) and the thin gold film (dashed lines). Given the mean values, we calculated the elastic modulus of PDMS with the Hertz model for a spherical rigid indenter in an infinite half-space obtaining the following values: $E_{0\%}^{PDMS} = 2.8 \pm 0.5\ MPa, E_{5\%}^{PDMS} = 2.6 \pm 0.5\ MPa, E_{10\%}^{PDMS} = 2.8 \pm 0.2\ MPa$. To calculate the elastic modulus of the gold thin film, we used the linear relationship between force and displacement as predicted by the indentation model of rigid thin film deposited on a compliant substrate.[24] The values obtained are: $E_{0\%}^{Au} = 92 \pm 23\ GPa, E_{5\%}^{Au} = 85 \pm 27\ GPa, E_{10\%}^{Au} = 125 \pm 43\ GPa$. The elastic modulus of the gold thin film is found to be comparable with the bulk value.

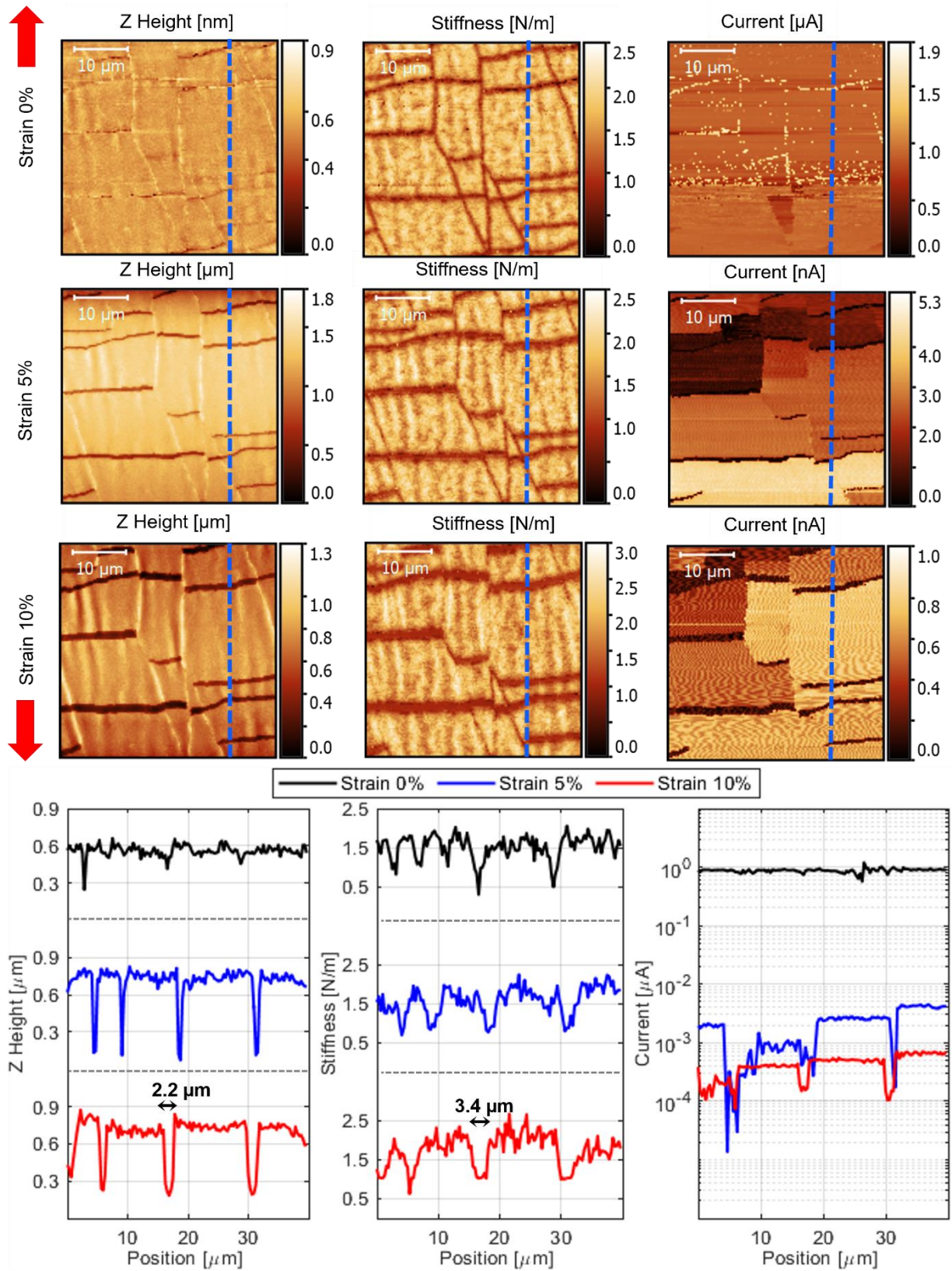

Figure 3. In-situ AFM multichannel acquisition. a) Morphology, stiffness, and current maps as a function of strain acquired on the same region of a microcracked gold film deposited on PDMS elastomer. The strain direction is represented by the two red arrows on the left. b) Height, stiffness, and current profiles extracted from the AFM maps. The dashed blue lines in a) indicate the positions of the profiles.

Current maps shown in **Figure 3** were obtained by applying a potential difference of 100 mV for strain 0% and 10V for greater strain values, between the sample and the tip. At the same time, we apply a voltage difference of 7V between the two ends of the freestanding film to monitor its macroscopic conductivity and to induce a lateral contrast in the current maps. At 0% strain, the gold film is entirely conductive, and the presence of the microcracks does not seem to have a major impact. In this regime the measured current is limited by the tip-sample contact. At higher strain values instead, there is a drop in the conductivity of the sample. The measured currents are three orders of magnitude lower. Close observation of the current maps and comparison with the height maps allows to identify the two crucial factors that impact on the local current value: First, the height map shows that the microcracks separate the gold thin film into individual fragments and on the current map we see that each of such fragments is characterized by constant current signal. Accordingly, we can conclude that a high conductivity is preserved within a fragment and the current is limited by how the fragment is connected to the rest of the film. Second, the map shows that different fragments are characterized by different current values and higher currents are observed in fragments positioned closer to the bottom of the map. This correlates with the potential gradient that builds up along the current transport path through the microcracked film driven by the external potentials applied to the sample ($V_{top} = 0$V, $V_{bottom} = 7$V, $V_{tip} = 0$V). Hence, the current is also controlled by the local potential that builds-up on individual gold fragments. Barriers in the current transport path due to weakly connected fragments are overcome by stronger local electric fields hence causing a larger potential step between fragments. The current map contains therefore also important information on how the current transport through the microcracked film evolves during strain.

To investigate the current transport onto individual fragments in more detail, we perform conductive AFM I-V scans at different strain values (**Figure 4b**). In the figure we compare the local conducting AFM analysis with the overall current flowing through the sample on normalized linear scale and logarithmic scale. Both, local AFM-current as well as macroscopic sample current, show a significant decrease with increasing strain. However, the macroscopic I-V curves maintain a linear, ohmic behavior, while the local current curves show a transition from linear at 0% strain to superlinear at elevated strains. The observed shape suggests that charge transport at the microscale between the gold fragments occurs by a field enhanced tunneling effect. Similar measurement curves have been obtained by studying the tunnel effect in gold nanogap junctions.[29,30] A quantitative description of tunneling transport across metal-insulator-metal systems is provided by the Simmons model.[31] The model introduces as parameters the insulator width ($s$), the height of the potential barrier ($\varphi$), and the overall current scale ($A$), which corresponds to the area of the two metal regions where charge transport by the tunneling effect occurs. The model provides a good fit to our data and the obtained parameter values are reported in **Table 1**. The barrier height of 5.4 eV and the gap size of a few angstroms indicates a dielectric mediated tunneling mechanism that proceeds through gaps between different gold fragments. Although the significant strain causes a widening of the cracks in the thin film, in the direction orthogonal to the strain, fragments remain closely spaced therefore enabling a tunneling mediated transfer path. In the macroscopic current measurement, the transition to the superlinear behavior is not present because a large number of fragments participate in the transport path. Accordingly, several small cracks have to be overcome and they all induce small potential steps driving field induced tunneling. At individual steps the potential drop remains strongly below the tunneling barrier height and a linear response is maintained.

|             | $s$ (Å)        | $\varphi$ (eV) | $A$ (Å²)       |
|:-----------:|:--------------:|:--------------:|:--------------:|
| Strain 5%   | $2.80 \pm 0.04$ | $5.7 \pm 0.2$  | $0.29 \pm 0.01$ |
| Strain 10 % | $2.86 \pm 0.02$ | $5.4 \pm 0.1$  | $0.11 \pm 0.01$ |

Table 1. Parameters of the Simmons model estimated by fitting the local I-V experimental data.

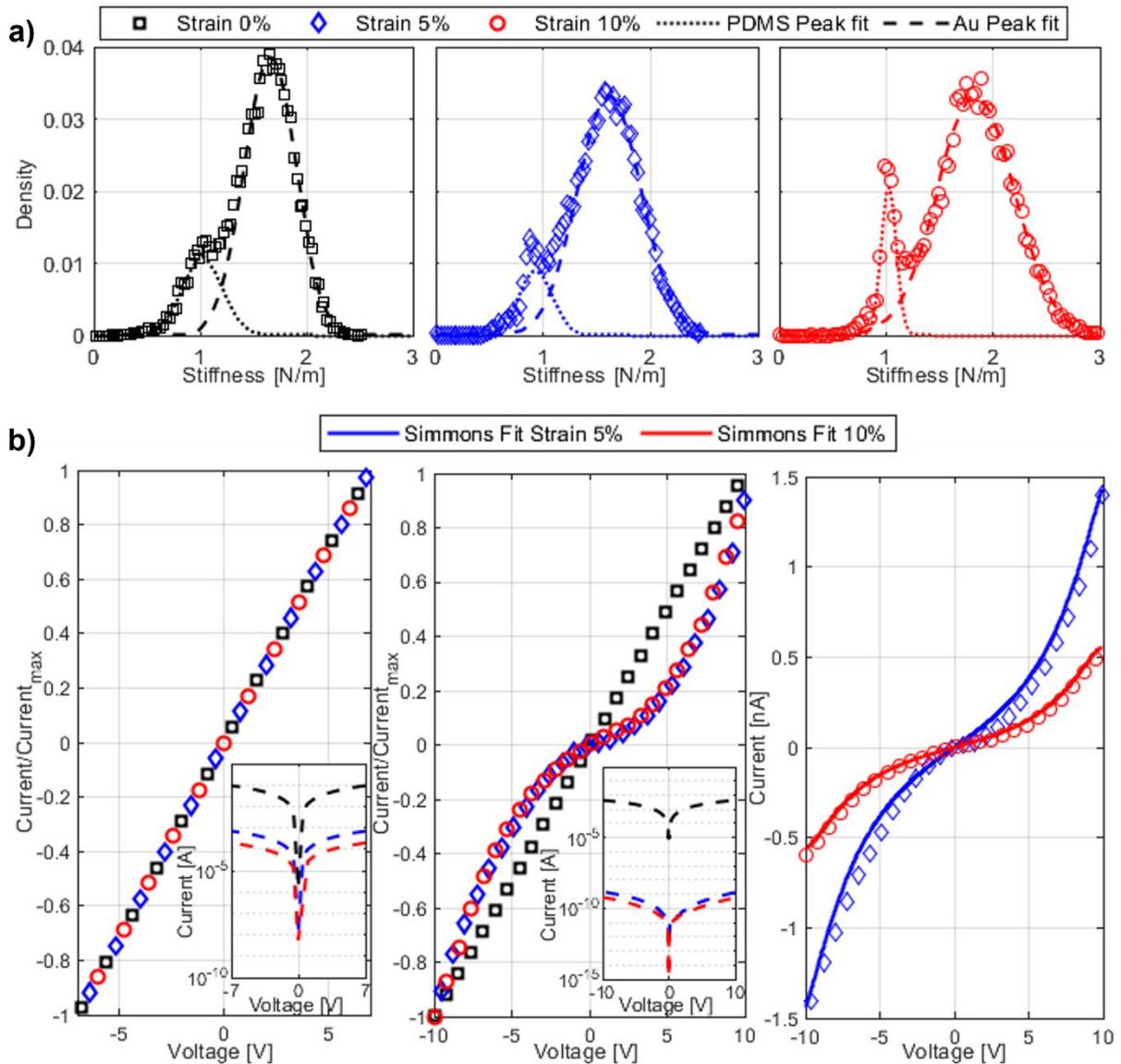

Figure 4 a) Stiffness histograms obtained from stiffness maps at different strains. The peak at lower stiffness corresponds to the PDMS, while the peak at higher stiffness corresponds to Au thin film. The dotted and dashed lines are Gaussian fit to estimate the average stiffness of PDMS and Au, respectively. b) the first graph on the left shows the macroscopic I-V curve of the entire metallic thin film at different strains. The graph in the middle shows the microscopic I-V curve of a single gold region acquired with C-AFM at different strains. The insets report the same data in a semilogarithmic plot to show the orders of magnitude. The graph on the right shows the microscopic I-V curve not normalized and fitted with the Simmons model describing the tunneling for a metal-insulator-metal system.

## 3. Conclusions and Discussion

Our work demonstrates a novel in-situ experimental method to investigate strain effects in materials and devices for stretchable electronics. Based on fast repetitive force spectroscopy acquisitions with a conductive AFM probe the method enables the microscopic investigation of morphological, mechanical, and electrical properties as a function of strain. The development became possible through a detailed investigation of possible artifacts that can occur when dynamic AFM techniques are performed on a free-standing substrate, only attached at its two ends to the clamps of a tensile stretcher. By deriving and testing the analytical equations that describe substrate deflection and resonant vibrations (**Equation 1** and **Equation 2**), we find the experimental conditions for stable, artifact-free image acquisitions. Disturbing sample deformation modes can be reduced by using elastic substrates with a sufficiently large thickness to length ratio, high elastic modulus or by operating force spectroscopy at slower approach and retract velocities to avoid excitation of resonant modes.

Once the conditions for stable measurements are met, our method provides an unprecedented multichannel imaging technique to correlate morphological defects generated during tensile strain with the mechanical and electrical response. As an example, we analyze the tensile deformation and conductivity changes of a thin gold film deposited with a chromium adhesion layer on silicon elastomer substrates (PDMS). With our experimental setup, we provide unique insight into the mechanisms of charge transport across gold regions separated by microcracks. For this the combination of all three imaging channels is crucial: First, the surface height mapping allows to identify gold fragments surrounded by microcracks and to investigate their morphological evolution during strain. Second, the micromechanical imaging channel provides quantitative measures on the local elastic modulus. It demonstrates that the externally applied strain is locally absorbed in micro-crack widening while gold fragments do not alter their physical extension and stiffness properties as increasing strain. Third, the local conducting properties demonstrate that individual gold fragments remain highly conductive, and transport is crucially determined by cracks separating the conducting fragments. We note that for gold on elastomer films, the processing conditions such as gold and adhesion layer thickness as well as pretreatment procedures have a crucial impact on how the microcrack pattern forms and separates islands. In the case studied here, we find at lower strains that a fully conductive, ohmic pathway remains present even though cracks widen in the direction perpendicular to the strain. Instead at higher strains, our detailed analysis of local IV-curves demonstrates the transition from the ohmic regime into a tunneling dominated regime, where local barriers have to be overcome by a field enhanced tunneling mechanism. We associate such barriers to microcracks oriented in the direction parallel to the strain. Only in this direction, cracks do not open significantly, thus leaving fragments in close enough proximity to permit tunneling transfer.

In conclusion, our work demonstrates a new in-situ experimental approach to investigate mechanical and electrical properties and their correlation at the microscale. It enables to map properties of stretched samples as a function of strain and to access experimentally the local electrical properties. Both are crucial aspects to understand and predict the properties of stretchable conductors. We highlight the value of the method by demonstrating the transition from local ohmic transport to field enhanced tunneling at increasing tensile strain in a microcracked gold layer. So far, models on the conductivity of thin films subjected to strain, assume ohmic transport in the defect-free parts of the metal layer, whereas through-thickness cracks are considered as completely isolating barriers.[17,18] However, our results show that charge transport occurs even if a conducting fragment is completely surrounded by microcracks due to the tunneling effect, introducing a conduction mechanism that has not been accounted for in models.

## 4. Experimental Section

*PDMS/Cr/Au preparation:* PDMS was obtained by mixing crosslinker and Sylgard 184 silicone in a ratio of 1:8. After intensive stirring, the mixture was degassed to remove air bubbles. Few µms of PAA were spin-coated on the glass substrate before casting PDMS to decrease the adhesion. After pouring the mixture, 20 minutes were waited to let it spread homogeneously onto the glass. Samples were then stored for one hour at 70 °C in an oven. Then the chromium adhesion layer (5 nm thickness) and gold (18 nm thickness) were deposited on the glass/PDMS substrates by thermal evaporation (source sample distance = 25 cm, vacuum pressure = $5.5 \, 10^{-6}$ mbar). Before clamping the sample in the strain stage, the sample was manually bent and twisted to pre-crack the gold surface to avoid the formation of a single crack cutting all conductive paths. The electrical contacts were made with copper tape and a conductive epoxy silver-based (without the hardener to keep it liquid), to increase the contact stability during strain variation.

*AFM Probe:* A Park System's NX10 AFM was used in the experiments. The Rocky Mountain Nanotechnology's probe 25Pt300B was used to perform fast repetitive force spectroscopy. The AFM tip used for resonant frequencies investigation was PPP CONTSCR, while for thickness measurement was NCHR, both from Nanosensors. Before each experiment, the tip sensitivity and force constant are calibrated by an indentation on a silicon surface and thermal tune method. The AFM tips 25Pt300B, PPP CONTSCR, and NCHR have spring constant equal to $18 \, N \, m^{-1}$, $0.2 \, N \, m^{-1}$, $5 \, N \, m^{-1}$, respectively.

*AFM resonant frequencies investigation*: A loudspeaker (Visaton K50) was used as a source of acoustic waves and controlled by a function generator (integrated in the Zurich Instruments MFLI Lock-in Amplifier). The vibrations of the sample were probed by a very soft AFM tip (PPP CONTSCR) in contact with a setpoint force of 10nN. The Position Sensitive Photo Detector voltage signal ($V_{a-b}$) was recorded by a Lock-in amplifier as a function of the loudspeaker excitation frequency. Since the impact of 20nm thick metal layer on the resonance frequencies is considered negligible, we performed the experiments on pure PDMS substrates (2.1mm thick). The strain stage shown in **Figure 1a** and **Figure S1b** was used to apply different tensile strain values and sample oscillation was measured in the frequency range from 10 Hz to 40 kHz.

*AFM multichannel imaging:* Height, Stiffness and current maps were generated by performing a fast force spectroscopy for each pixel of the image. Working in Contact mode, before the fast force spectroscopy the approach of AFM tip to the sample surface is performed. The approach is concluded when the setpoint force is reached. To move from pixel to pixel, the tip is lifted from the substrate. Then the XY stage is moved to place the new pixel below the tip and a tip approach is performed before the next force spectroscopy starts. As force spectroscopy parameters we set the maximum force to be 500nN, while the speed of the tip along the Z-axis was 100 $\mu m\ s^{-1}$. We set the lift height to be 1$\mu m$, the time for the pixel-to-pixel motion to be 5ms, and a pre-approach time of 100 $\mu s$. These parameters were optimized on PDMS/Cr/Au surfaces and an acquisition of a 128x128 pixels images of 40x40 $\mu m^2$ sample area is achieved in ca. 30 minutes. Current maps were obtained by applying a potential difference of 100 mV for strain 0% and 10V for greater strain values, between the sample and the tip. The macroscopic I-V curves were acquired with a SMU, tuning the voltage between -7 V and 7 V, and measuring the current.

**References**

[1]   M. J. Cordill, O. Glushko, J. Kreith, V. M. Marx, C. Kirchlechner, *Microelectron. Eng.* **2015**, *137*, DOI 10.1016/j.mee.2014.08.002.

[2]   M. Mohri, M. Nili-Ahmadabadi, M. PouryazdanPanah, H. Hahn, *Mater. Sci. Eng. A* **2016**, *668*, DOI 10.1016/j.msea.2016.05.044.

[3]   A. Kleinbichler, M. Bartosik, B. Völker, M. J. Cordill, *Adv. Eng. Mater.* **2017**, *19*, DOI 10.1002/adem.201600665.

[4]   U. Lang, T. Süss, N. Wojtas, J. Dual, *Exp. Mech.* **2010**, *50*, DOI 10.1007/s11340-009-9240-y.

[5]   X. Li, M. Sun, C. Shan, Q. Chen, X. Wei, *Adv. Mater. Interfaces* **2018**, *5*, DOI 10.1002/admi.201701246.

[6]   S. Roy, J. Ryan, S. Webster, D. Nepal, *Appl. Mech. Rev.* **2017**, *69*, DOI 10.1115/1.4038257.

[7]   M. A. Haque, M. T. A. Saif, *J. Mater. Res.* **2005**, *20*, DOI 10.1557/JMR.2005.0220.

[8]   F. Hang, D. Lu, R. J. Bailey, I. Jimenez-Palomar, U. Stachewicz, B. Cortes-Ballesteros, M. Davies, M. Zech, C. Bödefeld, A. H. Barber, *Nanotechnology* **2011**, *22*, DOI 10.1088/0957-4484/22/36/365708.

[9]   M. S. Bobji, B. Bhushan, *J. Mater. Res.* **2001**, *16*, DOI 10.1557/JMR.2001.0110.

[10]  P. Godard, P. O. Renault, D. Faurie, D. Thiaudière, *Appl. Phys. Lett.* **2017**, *110*, DOI 10.1063/1.4984135.

[11]  P. Godard, D. Faurie, P. O. Renault, *J. Appl. Phys.* **2020**, *127*, DOI 10.1063/1.5133715.

[12]  X. Li, W. Xu, M. A. Sutton, M. Mello, *IEEE Trans. Nanotechnol.* **2007**, *6*, DOI 10.1109/TNANO.2006.888527.

[13]  T. Cramer, L. Travaglini, S. Lai, L. Patruno, S. De Miranda, A. Bonfiglio, P. Cosseddu, B. Fraboni, *Sci. Rep.* **2016**, *6*, DOI 10.1038/srep38203.

[14]  A. Kleinbichler, M. J. Pfeifenberger, J. Zechner, N. R. Moody, D. F. Bahr, M. J. Cordill, *Jom* **2017**, *69*, DOI 10.1007/s11837-017-2496-2.

[15]  J. Zhong, D. He, *Sci. Rep.* **2015**, *5*, DOI 10.1038/srep12998.

[16]  P. O. Renault, P. Villain, C. Coupeau, P. Goudeau, K. F. Badawi, *Thin Solid Films* **2003**, *424*, DOI 10.1016/S0040-6090(02)01127-6.

[17]  O. Glushko, P. Kraker, M. J. Cordill, *Appl. Phys. Lett.* **2017**, *110*, DOI 10.1063/1.4982802.

[18]  O. Glushko, B. Putz, M. J. Cordill, *Thin Solid Films* **2020**, *699*, DOI 10.1016/j.tsf.2020.137906.

[19] N. Matsuhisa, X. Chen, Z. Bao, T. Someya, *Chem. Soc. Rev.* **2019**, *48*, DOI 10.1039/c8cs00814k.

[20] S. P. Lacour, S. Wagner, Z. Huang, Z. Suo, *Appl. Phys. Lett.* **2003**, *82*, DOI 10.1063/1.1565683.

[21] D. Qi, K. Zhang, G. Tian, B. Jiang, Y. Huang, *Adv. Mater.* **2020**, *33*, DOI 10.1002/adma.202003155.

[22] F. Decataldo, T. Cramer, D. Martelli, I. Gualandi, W. S. Korim, S. T. Yao, M. Tessarolo, M. Murgia, E. Scavetta, R. Amici, B. Fraboni, *Sci. Rep.* **2019**, *9*, DOI 10.1038/s41598-019-46967-2.

[23] S. P. Lacour, D. Chan, S. Wagner, T. Li, Z. Suo, *Appl. Phys. Lett.* **2006**, *88*, DOI 10.1063/1.2201874.

[24] G. Cortelli, L. Patruno, T. Cramer, M. Murgia, B. Fraboni, S. De Miranda, *ACS Appl. Nano Mater.* **2021**, *4*, DOI 10.1021/acsanm.1c01590.

[25] R. W. Clough, J. Penzien, *Dynamics of Structures*, **2002**.

[26] A. C. F.- Cripps, *Nanoindentation*, **2011**.

[27] O. Graudejus, T. Li, J. Cheng, N. Keiper, R. D. Ponce Wong, A. B. Pak, J. Abbas, *Appl. Phys. Lett.* **2017**, *110*, DOI 10.1063/1.4984207.

[28] D. Lee, J. R. Barber, M. D. Thouless, *Int. J. Eng. Sci.* **2009**, *47*, DOI 10.1016/j.ijengsci.2008.08.005.

[29] V. Dubois, S. N. Raja, P. Gehring, S. Caneva, H. S. J. van der Zant, F. Niklaus, G. Stemme, *Nat. Commun.* **2018**, *9*, DOI 10.1038/s41467-018-05785-2.

[30] A. Banerjee, S. U. H. Khan, S. Broadbent, R. Likhite, R. Looper, H. Kim, C. H. Mastrangelo, *Nanomaterials* **2019**, *9*, DOI 10.3390/nano9050727.

[31] J. G. Simmons, *J. Appl. Phys.* **1963**, *34*, DOI 10.1063/1.1702682.

**Supporting Information**

Supporting Information is available from the Wiley Online Library or from the author.

**Acknowledgements**

The authors gratefully acknowledge financial support from the EU Horizon 2020 FETOPEN-2018-2020 program (project "LION-HEARTED," grant agreement no. 828984).

Received: ((will be filled in by the editorial staff))
Revised: ((will be filled in by the editorial staff))
Published online: ((will be filled in by the editorial staff))

**Table of Content**

*Giorgio Cortelli, Luca Patruno, Tobias Cramer\*, Beatrice Fraboni, Stefano de Miranda\**

**In-situ force microscopy to investigate fracture in stretchable electronics: insights on local surface mechanics and conductivity**

We report an in-situ AFM technique to achieve multichannel imaging of height, surface conductivity and stiffness of a stretched, free-standing thin-film sample. As demonstrated for thin gold films deposited on a silicone membrane, the technique provides unique microscopic insight into strain induced defect formation and conductivity degradation, relevant to optimize materials properties of stretchable conductors.

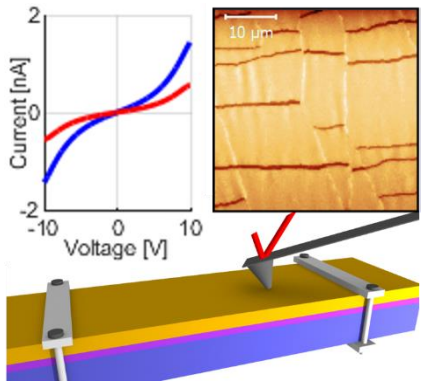

# Supporting Information

**In-situ force microscopy to investigate fracture in stretchable electronics: insights on local surface mechanics and conductivity**

*Giorgio Cortelli, Luca Patruno, Tobias Cramer\*, Beatrice Fraboni, Stefano de Miranda\**

The multichannel images of the in-situ AFM experimental method proposed in this work are obtained through fast repetition of the force spectroscopy experiment. Therefore, it is crucial to investigate the possible excitation of resonant oscillations of the free-standing sample that would interfere with the AFM acquisitions. A picture of the strain stage is shown in **Figure S1a**, while a picture of the experimental setup showing the free-standing configuration, is reported in **Figure S1b**. To understand the dependence of the resonant frequencies on the sample geometry and the strain, we developed a model starting from the well-known Rayleigh quotient method for vibration analysis and compared it with the experimental results.[25] The sample is modeled as a pre-stressed rectangular plate clamped at two ends. Geometrical variations due to large strains and Poisson effect have been accounted for. In fact, as the strain increases, the length increases, while the width and thickness decrease. Denoting with $u$ the transverse displacement of the plate and with $x$ the axis on the plate mid-surface and orthogonal to the clamps, we assumed $u(x) = \frac{1}{2}\left(1 - cos\left(\frac{2\pi x}{L}\right)\right)$ as approximating function for the first vibration mode of the plate (cylindrical bending), with $x \in [0, L]$. With these assumptions, using the Rayleigh quotient method, we obtained the following formula for the frequency of the first vibration mode:

$$f_0^{1st}(\varepsilon) = \frac{\pi h}{3L^2} \frac{(1-\nu\varepsilon)}{(1+\varepsilon)^2} \sqrt{\frac{E}{\rho(1-\nu^2)}\left(1 + \frac{3L^2}{\pi^2 h^2} \frac{\varepsilon(1-\nu^2)(1+\varepsilon)^2}{(1-\nu\varepsilon)^2}\right)} \quad (S1)$$

where $L$ and $h$ are the length and the thickness of the sample, respectively, while $\rho$ is the density, $E$ is the elastic modulus, $\nu$ is the Poisson's ratio and $\varepsilon$ is the strain.

We then investigated the impact on force-indentation curves of the bending of the sample. In fact, considering the force applied from the AFM tip as a concentrated force, it must be ensured that the bending of the free-standing sample can be neglected with respect to the indentation of the tip inside the sample. To get an estimate of the bending at the center of the sample, we assumed the plate to behave like a beam clamped on both ends. The stiffness of the beam is then given by the formula:

$$K_{flex} = \frac{192EI}{L^3(1-\nu^2)} \tag{S2}$$

where $I = \frac{Bh^3}{12}$ is the beam cross section inertia, and $B$ is the sample width. Notice that such approach is valid only if $B$ is smaller than $L$, and the error for $B = L$ is approximately 10 %. The plate deflection is then $\delta_{flex} = F/K_{flex}$. The validity of this description has been confirmed with numerical simulations. An estimate of the indentation of the AFM tip on the PDMS sample is given by the well-known Hertz model for a rigid spherical indenter in an infinite half-space. Therefore, it can be expressed as $\delta_{hertz} = \left(\frac{3F(1-\nu^2)}{4E\sqrt{R}}\right)^{2/3}$. The ratio of $\delta_{flex}$ and $\delta_{hertz}$ allows understanding of which between indentation and bending is the dominant term. The ratio can be expressed as:

$$\frac{\delta_{flex}}{\delta_{hertz}} = \frac{L^3}{192I}\left(\frac{8FR(1-\nu^2)}{9E}\right)^{1/3} \tag{S3}$$

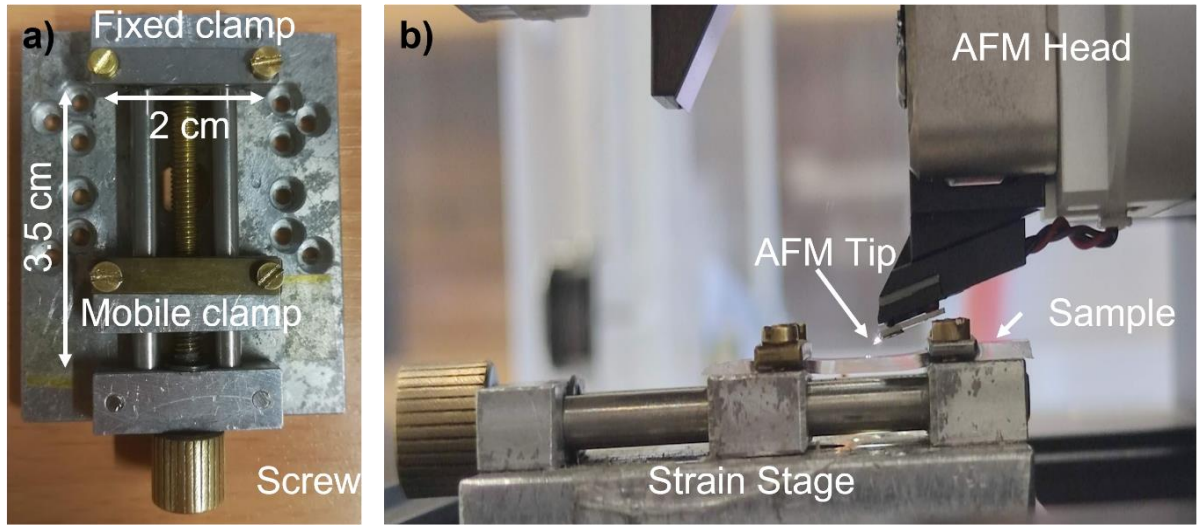

Figure S1. a) Photo of the strain stage with its dimensions. b) Photo of the experimental setup with sample clamped to the strain stage under the AFM tip.